\documentclass[a4paper,10pt,twoside]{article}
\usepackage[cp1251]{inputenc}
\usepackage[english,russian]{babel}
\usepackage{amsmath}
\usepackage{amssymb}
\usepackage{amsthm}
\usepackage{graphicx}

\theoremstyle{definition}

\theoremstyle{remark}

\newcommand{\bP}{{\bf P}}

\newcommand{\bX}{{\bf X}}

\newcommand{\bz}{{\bf z}}

\newcommand{\hX}{{\hat{\bf X}}}

\newcommand{\be}{\begin{equation}}
\newcommand{\ee}{\end{equation}}
\newcommand{\bay}{\begin{eqnarray}}
\newcommand{\eay}{\end{eqnarray}}
 \author{}
 \title{Model of the effective separable potential in the problem of three one-dimensional quantum particles}

\begin{document}
 \addtolength{\hoffset}{-3.1cm}
  \addtolength{\voffset}{-3cm}

\maketitle

\centerline{\large S.~B.~Levin$^{1,2}$, A.~S.~Bagmutov$^{2,3}$, and V.~O.~Toropov$^{1,2}$}

\vskip0.5cm

\centerline{$^1$ St.Petersburg State University}

\vskip0.5cm

\centerline{$^2$ Institute for Problems in Mechanical Engineering
of the Russian Academy of Sciences}

\vskip0.5cm

\centerline{$^3$ St.Petersburg Institute of Fine Mechanics and Optics}

\vskip1cm

\section{Problem formulation}

The goal of this paper is to construct an effective model
for studying the asymptotic solution of the scattering problem of
three one-dimensional quantum particles with finite
(short-range) attractive pair potentials. The asymptotic nature
of the solution is defined by the rapid decrease
in its discrepancy in the Schr\"odinger equation.
We consider the scattering problem of three one-dimensional
quantum particles of equal masses~($m_i=\frac12,\ i=1,2,3$).
The dynamics of the system is described by the Schr\"odinger equation
\be
H_{lab}\Psi^{lab}=E\Psi^{lab}  \ \ \ \ \ H_{lab}=-\Delta_{\bz}+V(\bz),\ \ \ \bz=(z_1,z_2,z_3),
\label{shr-0}
\ee
where~$z_j$ are the coordinates of particles in the laboratory frame of reference,
\be
V(\bz)=\mathop{\sum}\limits_{i<j;i,j=1,2,3} v_k(|z_i-z_j|), \ \ \
\bz\in \mathbb{R}^3,\ \ \ z_j\in \mathbb{R},\ \ j=1,2,3.
\label{eqc-0}
\ee
Here the indices~$(i,j,k)$ form an even permutation.

Let us pass to the reference system
associated with the center of mass of the three-particle system
$$
z_1+z_2+z_3=0,
$$
and introduce three systems of Jacobi coordinates~$(x_j,y_j),\ j=1,2,3$
on the resulting hyperplane~$S=\{\bz\in \mathbb{R}^3:\ z_1+z_2+z_3=0\}$;
each of the systems is associated with one of the pair subsystems.
All three coordinate systems are equivalent
and related to each other by rotation transformations.
Namely,
\be
x_1=\frac{1}{\sqrt{2}}(z_3-z_2),\ \ \ x_2=\frac{1}{\sqrt{2}}(z_1-z_3),\ \ \
x_3=\frac{1}{\sqrt{2}}(z_2-z_1),\ \
\label{yacobi}
\ee
$$
y_j=\sqrt{\frac{3}{2}}z_j,\ \ \ \ j=1,2,3.
$$
We also note the relations
connecting the pair Jacobi coordinates
with the chosen coordinate system
\be
x_2=-\frac{\sqrt{3}}{2}y_1-\frac12 x_1,\ \ \ \
x_3=\frac{\sqrt{3}}{2}y_1-\frac12 x_1.
\label{rot}
\ee

In the new notation, we arrive at the following Schr\"odinger equation for~$S$:
\be
H\Psi=E\Psi,\ \ \ \ H=-\Delta+V(\bX),\ \ \ V(\bX)=\mathop{\sum}\limits_{i=1}^3 v_i(x_i).
\label{ham-new}
\ee
We assume here that the operator~$\Delta$
on the hyperplane~$S$ is the Laplace--Beltrami operator
$$
\Delta=\frac{\partial^2}{\partial x^2}+\frac{\partial^2}{\partial y^2},
$$
invariant with respect to the choice of a~specific Jacobi coordinate system.
We assume that all pair potentials~$v_j(x_j),\ j=1,2,3$
are identical, finite, and are attractive potentials
supporting one bound state.
We also assume that the pair potential
is a~smooth function~$v_i\in C_{\mathbb{R}},\ i=1,2,3$,
although this condition can be weakened.

We note that the restrictions associated with the equality
of particle masses, as well as with the equality of pair potentials,
are not fundamental and are introduced only to technically simplify
the solution of the problem. The solution of the problem
allows a direct generalization to the case of arbitrary masses
and arbitrary finite pair potentials (both attractive and repulsive)
in different pairs.

We note that the scattering problem of three
one-dimensional quantum particles~$3\rightarrow 3$
(all three particles are free in the initial and final states of the system)
in the situation of repulsive pair potentials
was considered within the framework of the diffraction approach
in~\cite{gaudin}--\cite{BL-1}.
At the same time, the asymptotic eigenfunctions
of the Schr\"odinger operator were constructed.
The case of~$N$-particle scattering~\cite{BK} was also considered separately.
Later, in~\cite{BBL-1},
the limiting values of the resolvent kernel of the Schr\"odinger operator
were constructed on the continuous spectrum
within the framework of the alternating Schwartz method.
This, in turn, allowed isolating the asymptotics
of the three-particle eigenfunctions
of the absolutely continuous spectrum~\cite{BBL-2},
which confirmed the results obtained earlier
within the diffraction approach.

We emphasize that the analysis of scattering
in the system of three particles on a~straight line
both is the first step towards studying
the problem of three particles in three-dimensional space,
and is interesting in itself. This is confirmed by the fact
that the system of three (neutral or charged) particles
on a~line has been intensively studied
for many years (see, for example,~\cite{Lieb}--\cite{Chuka-2007}).
Currently, interest in such systems has increased
since they have been implemented experimentally
(see~\cite{Gorlitz}--\cite{Estive-2006}).
Therefore, mathematically correct and logically clear
numerical procedures for constructing the states
of the continuous spectrum of the Schr\"odinger operator
of three-particle systems in the one-dimensional case
are obviously in demand. The proposed approach
allows a natural generalization
to the case of slowly decreasing pair potentials
in one dimension and in higher dimensions.
A~generalization to the case of a~larger number of particles
is also possible in principle.

We note that an~attempt was undertaken in~\cite{jetp-2022}
to take into account
the presence in the system of attractive pair potentials
supporting the bound states,
within the framework of the diffraction approach,
for the~$2\rightarrow 2(3)$ scattering problem
(in the initial state, the system contains
a~free particle and a~two-particle cluster
in a~bound state;
in the final state, the system may consist of
both a~free particle and a~cluster in a~bound state,
and of three free particles)
in the case of three one-dimensional quantum particles.
Although the ideological line of the paper was correct,
certain weaknesses appeared in its implementation.
In this paper, the outline proposed in~\cite{jetp-2022}
is corrected due to a~more accurate description
of the ansatz of the three-particle solution
while preserving the main ideas proposed in~\cite{jetp-2022}.

Before we pass to the description
of the results obtained in this study,
let us recall the optical model that is well known in nuclear physics,
proposed by Feshbach, Porter, and Weisskopf in~1954~\cite{opt-model}
to describe the averaged behavior of cross sections.
The model got its name from the analogy between the scattering of particles
on a~kernel and the passage of light through a~semitransparent sphere.
In the optical model, it is assumed that the kernel can be described
by a~complex potential well
using the so-called optical potential
$$
U(r)=V_{opt}(r)+iW_{opt}(r),
$$
where the imaginary part~$W_{opt}(r)$ describes the absorption
of particles of the incident ray.
The imaginary part of the optical potential is proportional
to the average probability of transition
from a~single-particle state
(defined by the initial state of the incident particle)
to a~more complex state of the compound system
(a target kernel and a particle).
In other words, such a~phenomenological structure
of the potential in the optical model
takes into account the transition of the system
to a~new scattering channel (if there is one)
and is energy-dependent.

In this paper, we reduce the solution of the scattering problem
of a particle on a two-particle cluster
to the solution of some auxiliary boundary value problem
for the Schr\"odinger operator~$\hat{H}$
in a~circle of large radius
with a radiation condition at the boundary.
This boundary condition corresponds to the breakup of the system
into three free particles.
The potential in the Schr\"odinger operator in this case
is equal to the sum of the original three-particle potential~$V$
and some additional complex separable term~$V_{sep}$,
depending on the scattering energy.
All pair amplitudes of rearrangement of the two-particle clusters
are expressed in terms of the functionals of the solution~$\Phi$
of the resulting boundary value problem.
The amplitude of the breakup of the original system
into three free particles is also constructed using~$\Phi$.
The complete (numerical) solution of the original scattering problem
is also constructed in terms of the solution of the boundary value problem.
In view of the above,
the additional separable potential~$V_{sep}$ constructed in our model
makes sense and is similar (in structure and content)
to the optical potential mentioned above.
Namely, on one hand, the constructed model boundary value problem
contains an~asymptotic boundary condition
reflecting the breakup of the system into three free particles,
that is, the~$2\rightarrow 3$ scattering channel.
On the other hand, the structure of the operator
contains an~additional energy-dependent potential
that takes into account the possibility of reclustering,
that is, it takes into account
the presence of the asymptotic scattering channels~$2\rightarrow 2$.
In this sense, the constructed separable potential arises
as a~consequence of eliminating
the cluster scattering channels~$2\rightarrow 2$
while preserving the scattering channel~$2\rightarrow 3$.

\section{Main results}

1. In this paper, we propose a~method for calculating
the asymptotic eigenfunctions of the absolutely continuous spectrum
of the Schr\"odinger operator
for the~$2\rightarrow 2(3)$ scattering problem
of three one-dimensional quantum particles
with finite pair potentials supporting the bound states.

The proposed method allows reducing the asymptotic solution
of the original scattering problem to the construction
of some model inhomogeneous boundary value problem
for the ``extended'' Schr\"odinger operator
with a~potential equal to the sum
of the total initial potential~$V$
and some additional term. This additional term
is a~complex energy-dependent separable potential
of finite rank, similar in content to the so-called optical potential.
Thus, the asymptotic solution of the original scattering problem
is completely defined in terms of Green's function
of the constructed model inhomogeneous boundary value problem.

2. In this paper, we propose a~method
for solving the constructed model boundary value problem.
The method is based on the implementation
of the alternating Schwartz method~\cite{BBsin},
which is one of the variants of the Faddeev equation method.

3. To justify the applicability of the alternating method,
we give estimates for the second iteration
of the reflection operators~$\Gamma_i\Gamma_j,\ \ i\neq j$,
where~$\Gamma_i=V^{(i)}_{sep}R_i$.
Here~$V^{(i)}_{sep}$ is a~separable potential of rank one,~$R_i=(-\Delta+V+V^{(i)}_{sep}-E)^{-1}$.

4. We give an~estimate for the rate of decrease
of the~$n$th iteration of reflection operators.

5. In this paper, we present the results
of the numerical implementation of the proposed method
(see Sec.~7) for the technically simplest situation
of particles of equal masses and identical finite attractive pair potentials
supporting one bound state in each pair,
for two various values of the total energy
of the system~$E=0.5$ and~$E=2$.
For each of the energies, we calculate the reclustering amplitudes
in each pair scattering channel
and the breakup amplitude in the three-particle channel, and construct
a~numerically complete asymptotic solution
of the original scattering problem.
When constructing Green's function of the model boundary value problem,
we take into account the contributions
up to and including the second iteration of reflection operators.

6. We note that the method proposed in this paper
for an asymptotic solution of the~$2\rightarrow 2(3)$ scattering problem
can be easily generalized to the case of short-range pair potentials,
as well as to the~$3\rightarrow 2(3)$ scattering problem.
In principle, a~generalization to the case
of a~larger number of particles is possible.
We also note that the method proposed within the diffraction approach
for solving the three-body problem
in the presence of a~discrete spectrum
in pair subsystems complements the results,
including the numerical ones,
obtained earlier in~\cite{BL2}--\cite{BL3}
for the case of one-dimensional repulsive pair potentials.

\section{Scattering problem~$2\rightarrow 2(3)$ }

We consider here the~$2\rightarrow 2(3)$ scattering process
within the framework of the diffraction approach.
We assume that the pair potentials~$v_i,\ \ i=1,2,3$ are finite,
continuously differentiable (the smoothness requirement can be relaxed),
even, nonpositive, and supporting one bound state.
We rely here on the Calogero criterion~\cite{calodjero}
(and its generalization for the potentials set on the axis),
which defines the number of bound states in the two-body system.
We also assume that the pair~$(x,y)$ is a~pair of Jacobi coordinates
corresponding to the three-body system. Here~$x\in\mathbb{R},\ \ y\in\mathbb{R}$.
We assume that the masses of particles and the pair potentials are identical.

We will study the scattering problem~$2\rightarrow 2(3)$
of three particles on an~axis, that is,
the coordinate of each particle is characterized
by a~real number. More precisely, we will study
the scattering of a~bound pair on a~third particle,
using the formalism of the diffraction approach,
described in detail in~\cite{BMS1},~\cite{BMS2},~\cite{BL-1}.
Within the framework of this formalism,
the configuration space of the problem
after separating the dynamics of the center of mass
of the entire system is the plane~$S$,
each of the three pairs of Jacobi coordinates~$(x_j,y_j),\ j=1,2,3$
forms an~oriented coordinate system on~$S$,
these coordinate systems
(as well as the pairs of Jacobi coordinates themselves,
corresponding to two arbitrary various pair subsystems)
are related by a~rotation transformation.
The complete support of potential
(the union of three pair supports of potential)
is a family of three rays-``screens''~$\ l_j, j=1,2,3$
intersecting at one point with the vicinities.
In this case of finite pair potentials,
the support of the total potential
is the union of three oriented strips on the plane,
and the width of each strip in this case is defined
by the support of the corresponding pair potential.
Each of the ``screens'' with index~$j,\ j=1,2,3$
defines a~region in the configuration space
such that the particles in the pair~$j$ coincide,
that is, the equality~$x_j=0$ holds.
Thus, along the ``screen'' with index~$j$,
the Jacobi coordinate~$y_j$ changes,
and orthogonally to the screen,
with a~fixed orientation of the coordinate system,
the Jacobi coordinate~$x_j$ changes.
The sign of the coordinate~$x_j$ is defined
by the parity of the permutation of particles
in the pair~$j$, and the sign of the coordinate~$y_j$
is determined by the parity of the permutation
of particle~$j$ and the center of mass
of the pair of particles~$k$ and~$l$.
We assume here that the triple of indices~$(j,k,l)$
is formed by the permutation of numbers~$(1,2,3)$.

We also assume that the asymptotics
of the solution of the Schr\"odinger equation
$$
(H-E)\Psi=0,
$$
which almost everywhere satisfies the radiation conditions
at infinity in the configuration space
$$
\left(\frac{\partial}{\partial n}-i\sqrt{E}\right)\Psi|_{X=R}=O\left(|R|^{-3/2}\right),\ \ \ X=\sqrt{x^2+y^2},
$$
is arranged as follows:
\be
\Psi\sim  \psi_{in}(\bX,\bP) +
\mathop{\sum}\limits_{j=1}^3
\mathop{\sum}\limits_{\tau_j\in\{+,-\}}
a_j^{\tau_j}\psi_j^{\tau_j}(\bX,\bP) + A(\hX,\bP)\frac{e^{i\sqrt{E}X}}{\sqrt{X}} + O\left(\frac{1}{X^{3/2}}\right).
\label{ass-ggg}
\ee
Here the differentiation operator~$\frac{\partial}{\partial n}$
denotes the differentiation operator along the normal
to the boundary of the region,
a~circle of large radius~$R$.
We also use the notation
$$
\psi_{in}(\bX,\bP)\equiv e^{-ip_1y_1}\varphi_1^-(x_1),
$$
\be
\psi_j^{\tau_j}(\bX,\bP)\equiv e^{ip_jy_j}\varphi_j^{\tau_j}(x_j).
\label{psi-ggg}
\ee
The index~$\tau_j=\pm$ defines the sign of the half-screen~$l_j$
on which the function~$\psi_j^{\tau_j}$ is defined,
or, in other words, it defines the sign of the variable~$y_j$.
The notation~$\varphi$ stands for the pair bound state.

We also use the notation
$$
\bX=\left(
\begin{tabular}{c}
$x$ \\
$y$ \\
\end{tabular}
\right)\in \mathbb{R}^2,\ \ \ \ \bP=\left(
\begin{tabular}{c}
$k$ \\
$p$ \\
\end{tabular}
\right)\in \mathbb{R}^2,\ \ \ \ \hX=\frac{\bX}{X}.
$$
Here the symbols~$k\in\mathbb{R}$ and~$p\in\mathbb{R}$ stand for the description of the moments
that are conjugate in a Fourier sense to the Jacobi coordinates~$x$ and~$y$.

We assume that the particles of the pair~$j=1$
in the initial state are in a~bound state~$\varphi_1^-$
with energy~$\kappa_1<0$.
The functions~$\varphi_j^{\tau_j}$ satisfy the normalization condition
\be
\int_\mathbb{R}|\varphi_j^{\tau_j}(x)|^2 dx=1,\ \ \ \ j=1,2,3,\ \ \ \ \tau_j\in\{+,-\}.
\label{norm-ggg}
\ee
In this case, the cluster pair solution~$\varphi_j^{\tau_j}$
is an~even function on the support
of the finite pair potential~$v(x_j)$
and decays exponentially outside the support of potential
as the value of~$|x_j|$ grows.

We note that the first term in the expression~(\ref{ass-ggg})
corresponds to the incident wave.
In this case, the following relations are satisfied:
$$
y_1<0,\ \ p_1>0.
$$
The second term in the expression~(\ref{ass-ggg})
corresponds to the superposition of outgoing cluster waves
(the~$2\rightarrow 2$ processes) with amplitudes~$a_j^{\tau_j}$.
Here the index~$j=1,2,3$ denotes the number of the pair subsystem,
and the index~$\tau\in \{+,-\}$ defines the parity
of permutation of the coordinate of the particle~$j$
and the center of mass of the subsystem with index~$j$.
In other words, the index~$\tau_j$ corresponds to the sign
of the Jacobi coordinate~$y_j$ and therefore
defines the ``half-screen''~$\ l_j^{\tau_j}$.
For the outgoing cluster waves, the following relations hold:
$$
y_j>0,\ \ \ p_j>0\ \ \ \ \mbox{or}\ \ \ \ y_j<0,\ \ \ p_j<0.
$$
We note that each diverging wave with amplitude~$a_j^{\tau_j}$
is defined only on the half-screen that corresponds
to the index~$\tau_j$. On the half-screen with index~$-\tau_j$,
it continues by zero.
Finally, the third term in the expression~(\ref{ass-ggg})
corresponds to the~$2\rightarrow 3$ decay process
and describes the diverging circular wave with amplitude~$A(\hX,\bP)$.

Let us now rearrange the solution
of the original scattering problem in a~bounded,
but sufficiently large for the asymptotics of the solution to occur,
region.
The purpose of such a~rearrangement, or “deformation”, of the solution
of the original problem is to construct a~functional connection
between the asymptotic scattering channels~$2\rightarrow 2$ and~$2\rightarrow 3$.
Establishing such a~connection allows effective eliminating of
the cluster scattering channels and formulating a~new model boundary value problem
for the ``deformed'' part of the solution
in the~$2\rightarrow 3$ scattering channel.
In terms of the solution of such a~model problem,
we reconstruct the complete asymptotic solution
of the original scattering problem~$2\rightarrow 2(3)$.
Let us now describe the procedure
for the rearrangement of the solution more precisely.
We are going to construct a~set of equations
relating the amplitudes~$a_j^{\tau_j},\ j=1,2,3;\ \tau_j=\pm$
of the~$2\rightarrow 2$ scattering processes and the ``deformed''
in a~bounded region part of the solution
corresponding to the breakup process~$2\rightarrow 3$.
In this case, we know the solution of the Schr\"odinger equation
only in the asymptotic region of the configuration space for~$X\gg 1$.
Let us introduce a~smooth cut-off function that “cuts off” the~$\ $ solution
for limiting and small values of~$X$.
Multiplying the exact solution of the scattering problem
(or some part of it) by such a~cut-off function,
we obtain a~new function that remains an~exact
(up to the terms of the next order of smallness)
solution of the Schr\"odinger equation
for large~$X$, and for limiting and small values of~$X $,
although it will no longer be an~exact solution of the Schr\"odinger equation,
will generate a~known nonzero discrepancy
in a~bounded region of the configuration space.

This fact allows implementing the following outline for solving the scattering problem.
At the first stage, we use Green's second formula
on the plane and find, albeit in terms of some cut-off function,
the connection between the scattering amplitudes of the~$2\rightarrow 2$ processes
and the functionals of the ``deformed'' in a~bounded region
of the hyperplane~$S$ part of the solution, corresponding to the process~$2\rightarrow 3$.
In other words, we connect the cluster solutions of the scattering problem
(corresponding to the~$2\rightarrow 2$ processes)
with a~``deformed'' diverging circular wave
(corresponding to the~$2\rightarrow 3$ processes).
We hence exclude the two-particle scattering channels from consideration,
using their connection with a three-particle channel.
Such effective elimination of the interaction channels
in a~multichannel system always leads to the appearance
of some so-called optical potential, which is what we observe in this problem.

At the second stage of the solution of the scattering problem,
we construct an~inhomogeneous boundary value problem
for the part of the solution
which complements the deformed set of cluster solutions
to the complete asymptotic solution of the original scattering problem.
The asymptotics of this unknown part of the solution
at large distances behaves as a~diverging circular wave
with a~smooth amplitude and satisfies
the radiation conditions almost everywhere at infinity.

\section{Construction of the connections between cluster solutions and\\ a~``deformed'' diverging circular wave}

We begin the construction of the boundary value problem
by describing a~convenient representation
for an exact solution of the scattering problem.
Since the asymptotics of the solution at large distances
is known and has the form~(\ref{ass-ggg}),
we will seek a~solution in a~circle of large radius~$R\gg 1$
in the form of the sum of an~incident wave,
smoothly ``cut off'' for limiting and small values
of the hyperradius~$X$, cluster waves, also smoothly ``cut off''
for limiting and small values of the hyperradius~$X$,
and the unknown function~$\Phi$.
In this case, the function~$\Phi$ is defined everywhere
in a~circle of radius~$R$ and, as follows from the asymptotics~(\ref{ass-ggg}),
behaves at large values of~$X$ as a~diverging wave
with a~smooth amplitude. Looking ahead, we will say
that it is for the function~$\Phi$
that the boundary value problem will be constructed.

Let us introduce the radial cut-off function~$\zeta(X)\in C^2_{[0,\infty)}$ as follows:
\be
\zeta(X)=\left\{
\begin{tabular}{l}
$0,\ \ 0\le X<R_1,\ \ R_1\gg 1,$\\
$\mbox{monotonically increases from~0 to~1},\ \ R_1<X<R_2,$\\
$1,\ \ \ X>R_2.$\\
\end{tabular}
\right.
\label{srez-ggg}
\ee
We seek a~solution of the scattering problem in the following form:
\be
\Psi(\bX,\bP) =  \psi_{in}(\bX,\bP)\zeta(X) +
\mathop{\sum}\limits_{j=1}^3
\mathop{\sum}\limits_{\tau_j\in\{+,-\}}
a_j^{\tau_j}\psi_j^{\tau_j}(\bX,\bP)\zeta(X) + \Phi(\bX,\bP).
\label{exact-ggg}
\ee
Let us also introduce the notation
\be
\tilde{\psi}_{in}(\bX,\bP)\equiv \psi_{in}(\bX,\bP)\zeta(X),
\label{tild-in-ggg}
\ee
\be
\tilde{\psi}_j^{\tau_j}(\bX,\bP)\equiv \psi_j^{\tau_j}(\bX,\bP)\zeta(X).
\label{tild-ggg}
\ee

Let us now use Green's second formula
to construct a~connection between the cluster amplitude~$a_j^{\tau_j}$
and the functional of~$\Phi$. We need a~couple of equations
\be
\left\{
\begin{tabular}{l}
$(H-E)\Psi^*=0$ \\
$(H-E)\tilde{\psi}_j^{\tau_j}=-Q_j^{\tau_j}$,\\
\end{tabular}
\right.
\label{syst-ggg}
\ee
where the symbol~$*$ denotes complex conjugation.
We note that the discrepancy~$Q_j^{\tau_j}$ of the function~$\tilde{\psi}_j^{\tau_j}$
in the Schr\"odinger equation obviously differs from zero
where the cut-off function smoothly varies from~0 to~1,
and vanishes where the cut-off function is a~constant.
Since~$\tilde{\psi}_j^{\tau_j}$,
according to~(\ref{tild-ggg}) and~(\ref{psi-ggg}),
contains a~pair solution~$\varphi_j^{\tau_j}(x_j)$,
exponentially decreasing in~$|x_j|$
outside the support of the pair potential,
the residual~$Q_j^{\tau_j}$ is localized
in a curvilinear strip.
By choosing the values of~$R_1,\ R_2$ sufficiently large,
we can approximate the~given strip arbitrary well to the rectangular strip
\be
\Pi_j^{\tau_j}: \ R_1<|y_j|<R_2,
\label{def-pi}
\ee
the region of variation of~$|x_j|$ is defined
by the rate of exponential decrease
of the function~$\varphi_j^{\tau_j}$.
For definiteness, we assume that
$$
-d < x_j < d,\ \ \ \ |\varphi(d)|=O\left(\frac{1}{R}\right).
$$
The sign of the variable~$y_j$ is defined
by the value of the index~$\tau_j$, or, in other words,
by the choice of the half-screen
on which the corresponding cluster solution is defined.

Let us multiply the first of the equations of system~(\ref{syst-ggg})
by~$\tilde{\psi}_j^{\tau_j}$, and the second by~$\Psi^*$.
Subtract the second equation from the first
and integrate the result in a~circle~$B_R$ of large radius~$R$.

Applying Green's second formula, we arrive at the equation
\be
\mathop{\int}\limits_{\partial B_R}\left(\frac{\partial \tilde{\psi}_j^{\tau_j}}{\partial n}\Psi^*-
\frac{\partial \Psi^*}{\partial n}\tilde{\psi}_j^{\tau_j}\right)dl=
\mathop{\int}\limits_{B_R}Q_j^{\tau_j}\Psi^*d\sigma.
\label{green-ggg}
\ee
This equation relates the energy flows
carried by the distorted diverging wave~$\Phi$
and the distorted cluster wave~$\tilde{\psi}_j^{\tau_j}$.

We begin by considering the case~$\{j,\tau_j\}\neq \{1,-\}$.

\subsection{Case~$\{j,\tau_j\}\neq \{1,-\}$}

In this case, associating the Jacobi coordinate system
to the corresponding half-screen~$l_j^{\tau_j}$
and omitting the indices, we write Eq.~(\ref{green-ggg}) as
$$
{a_j^{\tau_j}}^*\left\{2ip\int_{-d}^d\varphi^2(x)-\int_{\Pi_j^{\tau_j}}d\bX Q_j^{\tau_j}(\bX,\bP)\varphi(x)e^{-ipy}\zeta(y)\right\}+
\int_{-d}^d dx \varphi(x)\left[ \Phi^*(\bX,\bP)ipe^{ipR}-e^{ipR}\frac{\partial}{\partial y}\Phi^*(\bX,\bP)\right]|_{y=R}=
$$
\be
=\int_{\Pi_j^{\tau_j}}d\bX Q_j^{\tau_j}(\bX,\bP)\Phi^*(\bX,\bP).
\label{green-1-ggg}
\ee
Here, in the case~$\tau_j=-$, the integral over~$dy$
is reduced to the integral over~$d|y|$.
Using the expression for the residual~$Q_j^{\tau_j}$,
\be
Q_j^{\tau_j}(\bX,\bP)=2ipe^{ipy}\zeta'(y)\varphi(x)+e^{ipy}\zeta''(y)\varphi(x),
\label{discrep-ggg}
\ee
we calculate the second term in curly brackets,
$$
J \equiv \int_{\Pi_j^{\tau_j}}d\bX Q_j^{\tau_j}(\bX,\bP)\varphi(x)e^{-ipy}\zeta(y) =
\int_{\Pi_j^{\tau_j}}dxdy\left[2ip\varphi^2(x)\zeta'(y)\zeta(y)+\varphi^2(x)\zeta''(y)\zeta(y)\right].
$$
The index~$\tau_j$, as above,
defines the half-strip over which the integral is taken.
Using the normalization condition~(\ref{norm-ggg})
of the cluster solution~$\varphi$ and its real nature,
as well as the properties of the cut-off function, we finally obtain
\be
J=2ip\int_{R_1}^{R_2}dy\left[\zeta^2(y)\right]'\frac12-\int_{R_1}^{R_2}dy\left[\zeta'(y)\right]^2=
ip-\alpha, \ \ \ \alpha\equiv \int_{R_1}^{R_2}dy\left[\zeta'(y)\right]^2 > 0.
\label{pr-ggg}
\ee
Finally, choosing the outer radius~$R$
of the region~$B_R$ sufficiently large
and fixing in this case the cutting parameters~$R_1,\ R_2$,
we note that the expression in square brackets
in Eq.~(\ref{green-1-ggg}) for~$y =R$ is of order~$O(\frac{1}{\sqrt{R}})$.
This follows directly from the structure of the asymptotics
of the function~$\Phi$, which was discussed above.
Neglecting this term in Eq.~(\ref{green-1-ggg}), we obtain
\be
a_j^{\tau_j}=\frac{1}{-ip+\alpha}\int_{\Pi_j^{\tau_j}}d\bX {Q_j^{\tau_j}}^*(\bX,\bP)\Phi(\bX,\bP)+
O\left(\frac{1}{\sqrt{R}}\right) =
\frac{1}{-ip+\alpha}\left<Q_j^{\tau_j}|\Phi\right>+O\left(\frac{1}{\sqrt{R}}\right).
\label{koeff-ggg}
\ee

Let us now pass to the description of the situation~$\{j,\tau_j\}= \{1,-\}$.

\subsection{Case~$\{j,\tau_j\}= \{1,-\}$}

In this case, the expression~(\ref{green-ggg})
generates an~integral over the half-strip~$\Pi_1^-$.
This means that the incident wave~$\tilde{\psi}_{in}$ (\ref{tild-in-ggg})
also contributes to the integral.
The expression~(\ref{green-1-ggg}) in this case must be modified to
$$
{a_1^-}^*\left\{2ip\int_{-d}^d\varphi^2(x)-\int_{\Pi_1^{-}}d\bX Q_1^-(\bX,\bP)\varphi(x)e^{-ipy}\zeta(y)\right\}+
\int_{-d}^d dx \varphi(x)\left[ \Phi^*(\bX,\bP)ipe^{ipR}-e^{ipR}\frac{\partial}{\partial y}\Phi^*(\bX,\bP)\right]|_{y=R}=
$$
\be
=\int_{\Pi_1^{-}}d\bX Q_1^-(\bX,\bP)\Phi^*(\bX,\bP)+\int_{\Pi_1^{-}}d\bX Q_1^-(\bX,\bP)\tilde{\psi}_{in}^*(\bX,\bP).
\label{green-1-in-ggg}
\ee
Repeating the calculations carried out above,
we arrive at the following equation:
\be
a_1^{-}=\frac{1}{-ip+\alpha}\left[\int_{\Pi_1^{-}}d\bX {Q_1^{-}}^*(\bX,\bP)\Phi(\bX,\bP)+
\int_{\Pi_1^{-}}d\bX {Q_1^{-}}^*(\bX,\bP)\tilde{\psi}_{in}(\bX,\bP)\right] + O\left(\frac{1}{\sqrt{R}}\right)  =
\label{koeff-in-ggg}
\ee
$$
=\frac{1}{-ip+\alpha}\left\{
\left<Q_1^{-}|\Phi\right> + \left<Q_1^{-}|\tilde{\psi}_{in}\right>\right\} + O\left(\frac{1}{\sqrt{R}}\right).
$$

Thus, Eqs.~(\ref{koeff-ggg}) and~(\ref{koeff-in-ggg}) describe the connections
between the cluster amplitudes~$a_j^{\tau_j}$
and the functionals of the ``deformed'' diverging circular wave~$\Phi$.
We note also that
the correction in the expressions~(\ref{koeff-ggg}) and~(\ref{koeff-in-ggg})
vanishes in the limit as~$R\rightarrow \infty$.

\section{Construction of the boundary value problem}

Let us pass to the construction of the boundary value problem
for the unknown function~$\Phi$.
For this purpose, let us substitute the expressions
obtained in~(\ref{koeff-ggg}) and~(\ref{koeff-in-ggg})
for the cluster amplitudes~$a_j^{\tau_j}$
into the representation~(\ref{exact-ggg})
of the solution of the scattering problem~$2\rightarrow 2(3)$,
\be
 \Psi(\bX,\bP) =  \tilde{\psi}_{in}(\bX,\bP) + \frac{1}{-ip+\alpha}
\mathop{\sum}\limits_{j=1}^3
\mathop{\sum}\limits_{\tau_j\in\{+,-\}}
\tilde{\psi}_j^{\tau_j}(\bX,\bP)\left<Q_j^{\tau_j}|\Phi\right> +
\frac{1}{-ip+\alpha}\tilde{\psi}_1^{-}(\bX,\bP)\left<Q_1^{-}|\tilde{\psi}_{in}\right> +
\Phi(\bX,\bP).
\label{exact-gran-ggg}
\ee

Let us act on Eq.~(\ref{exact-gran-ggg}) from the left and from the right
with the operator~$H-E$. As a result, we arrive at the equation
\be
(H-E)\Phi - \frac{1}{-ip+\alpha}
\mathop{\sum}\limits_{j=1}^3
\mathop{\sum}\limits_{\tau_j\in\{+,-\}}
Q_j^{\tau_j}\left<Q_j^{\tau_j}|\Phi\right> =  Q_1^{in} +
\frac{1}{-ip+\alpha}Q_1^{-}\left<Q_1^{-}|\tilde{\psi}_{in}\right>.
\label{prob-1-ggg}
\ee
The definition of the residual~$Q_j^{\tau_j}$ was given above
in Eq.~(\ref{syst-ggg}). We also use the notation
$Q_1^{in}$,
$$
(H-E)\tilde{\psi}_{in}=-Q_1^{in}.
$$

Let us now formulate the boundary value problem
for the function~$\Phi$ in a~circle~$B_R$ of large radius~$R$
with the radiation condition on the boundary~${\partial B_R}$,
\be
\left\{
\begin{tabular}{l}
$(\hat{H}-E)\Phi=Q_b$ \\
$\left(\frac{\partial \Phi}{\partial n} - i\sqrt{E}\Phi\right)|_{\partial B_R}= O\left(\frac{1}{R^{3/2}}\right)$.\\
\end{tabular}
\right.
\label{gran-prob-ggg}
\ee
We use here the notation
\be
\hat{H} \equiv H + V_{sep},\ \ \ \ Q_b \equiv Q_1^{in} +
\frac{1}{-ip+\alpha}Q_1^{-}\left<Q_1^{-}|\tilde{\psi}_{in}\right>.
\label{shr-new-ggg}
\ee
The Schr\"odinger operator~$H$ was defined in Eq.~(\ref{ham-new}),
and the separable potential of rank six~$V_{sep}$
is defined according to Eq.~(\ref{prob-1-ggg}),
\be
V_{sep} \equiv -\frac{1}{-ip+\alpha}
\mathop{\sum}\limits_{j=1}^3
\mathop{\sum}\limits_{\tau_j\in\{+,-\}}\left.
|Q_j^{\tau_j}\right>\left<Q_j^{\tau_j}|\right..
\label{separ-ggg}
\ee
We note here that all functions~$Q_b,\ Q_j^{\tau_j}$,
localized in the regions~$\Pi_j^{\tau_j}$~(\ref{def-pi})
on the hyperplane~$S$, are smooth in both variables
with the exception of the boundaries of the regions~$\Pi_j^{\tau_j}$,
on which, however, continuity is preserved.

The solution of the~$2\rightarrow 2(3)$ scattering problem
is reconstructed in terms of the solution
of the boundary value problem~(\ref{gran-prob-ggg})
using Eq.~(\ref{exact-gran-ggg}).

The construction itself of the solution
of the boundary value problem formulated in Eq.~(\ref{gran-prob-ggg})
to describe the~$2\rightarrow 2(3)$ scattering processes
in the system of three one-dimensional quantum particles
with finite attractive pair potentials
within the framework of the diffraction approach
is the next question to be discussed.

\section{Construction of the solution of the boundary value problem}
\subsection{Resolvent of the Schr\"odinger operator with a perturbation of rank one}

Before we pass to the discussion
of the boundary value problem (\ref{gran-prob-ggg})--(\ref{separ-ggg}) itself
with an~operator~$H$ containing a~separable perturbation of rank six~(\ref{separ-ggg}),
we consider a~simpler problem for an~operator~$H_1$
with a~separable perturbation of rank one~\cite{Simon}.
Consider the Schr\"odinger operator
$$
H_1=H_0+\beta|Q><Q|,\ \ \ \ \beta\in \mathbb{C}.
$$
The resolvent identity in this case takes the form
\be
(H_1-E)^{-1}-(H_0-E)^{-1}=-\beta (H_1-E)^{-1}|Q><Q|(H_0-E)^{-1}.
\label{resol-1}
\ee
Projecting Eq.~(\ref{resol-1}) from the left and right
onto the state~$Q$ and introducing the functions
$$
F_1(E)\equiv <Q|(H_1-E)^{-1}|Q>,\ \ \ \ \ F_0(E)\equiv <Q|(H_0-E)^{-1}|Q>,
$$
we obtain
$$
F_1(E)-F_0(E)=-\beta F_1(E)F_0(E).
$$
We arrive in this case at the connection equation
\be
F_1(E)=\frac{F_0(E)}{1+\beta F_0(E)}.
\label{connect-0}
\ee
Let us now rewrite the resolvent identity as
\be
(H_1-E)^{-1}-(H_0-E)^{-1}=-\beta (H_0-E)^{-1}|Q><Q|(H_1-E)^{-1},
\label{resol-2}
\ee
and project the resulting equation from the right onto the state $Q$,
$$
(H_1-E)^{-1}|Q>=(H_0-E)^{-1}|Q>-\beta F_1(E)(H_0-E)^{-1}|Q>.
$$
Using the connection condition~(\ref{connect-0}),
we finally obtain
\be
(H_1-E)^{-1}|Q>=\frac{1}{1+\beta F_0(E)}(H_0-E)^{-1}|Q>.
\label{project-0}
\ee
Substituting the resulting expression
into the resolvent identity~(\ref{resol-1}), we obtain
\be
(H_1-E)^{-1}-(H_0-E)^{-1}=-\frac{\beta}{1+\beta F_0(E)}(H_0-E)^{-1}|Q><Q|(H_0-E)^{-1}.
\label{new-old}
\ee
The resulting relation connects the resolvent
of the Schr\"odinger operator~$H_1$
with a~separable perturbation of rank one
with the resolvent of the unperturbed operator~$H_0$.

We now return to the original problem
and consider the resolvent of the Schr\"odinger operator~$H$
with a~perturbation~(\ref{separ-ggg}), that is,
the resolvent of an~operator
with a~separable perturbation of rank six.

\subsection{Separable perturbation of rank six~$V_{sep}$}

We now consider the resolvent
of the Schr\"odinger operator~$H$
with a~separable perturbation of rank six~$V_{sep}$~(\ref{separ-ggg})
and the unperturbed operator~$H_0$ defined in Eq.~(\ref{ham-new}),
\be
H=H_0+V_{sep},
\label{shred-6}
\ee
where
$$
H_0=-\frac{\partial^2}{\partial x^2} - \frac{\partial^2}{\partial y^2} +
\mathop{\sum}\limits_{j=1}^3 v_j(x_j),\ \ \ \ \ \
V_{sep} \equiv \beta
\mathop{\sum}\limits_{j=1}^3
\mathop{\sum}\limits_{\tau_j\in\{+,-\}}\left.
|Q_j^{\tau_j}\right>\left<Q_j^{\tau_j}|\right..
$$
Here we use the notation
\be
\beta\equiv  -\frac{1}{-ip+\alpha}.
\label{beta-def}
\ee

We note that the solution of the inhomogeneous boundary value problem of the form~(\ref{gran-prob-ggg}),
\be
\left\{
\begin{tabular}{l}
$({H}-E)\Phi=Q_b$ \\
$\left(\frac{\partial \Phi}{\partial n} - i\sqrt{E}\Phi\right)|_{\partial B_R}= O\left(\frac{1}{R^{3/2}}\right)$,\\
\end{tabular}
\right.
\label{gran-prob-h0}
\ee
with an~unperturbed Schr\"odinger operator~$H=H_0$
(albeit with repulsive pair potentials~$v_j,\ \ j=1,2,3$)
and localized inhomogeneity was in fact implemented
in~\cite{BL2}--\cite{BL3}. In the case of a~separable perturbation of rank one,
$$
H=H_0+V^{(1)}_{sep},\ \ \ \ \ V^{(1)}_{sep}=\beta|Q><Q|,
$$
the solution of the complete problem is given
in terms of the solution of the boundary value problem
for the unperturbed operator~$(H_0-E)^{-1}|Q_b>$ using
Eqs.~(\ref{new-old}),
$$
(H_1-E)^{-1}|Q_b>=(H_0-E)^{-1}|Q_b>-\frac{\beta}{1+\beta F_0(E)}<Q|(H_0-E)^{-1}|Q_b>(H_0-E)^{-1}|Q>.
$$

In the more general case of a~perturbation
representing the sum of a~certain number (six) of
separable potentials of rank one,
\be
V_{sep}=\mathop{\sum}\limits_{i=1}^6 V_{sep}^{(i)},\ \ \ \ \ \  V_{sep}^{(i)}\equiv \beta |Q_i><Q_i|,
\label{sep-sh}
\ee
we use the alternating Schwartz method,
which was developed in~\cite{BBsin} and later applied
to study the scattering problem of three one-dimensional quantum particles
in~\cite{BBL-1}--\cite{BBL-2}.
We note that the alternating Schwarz method
is some variant of the well-known Faddeev equation method.
To technically simplify the notation, we introduce here the redefinition
$$
V_{sep}^{(i)}\equiv \beta |Q_i><Q_i|=\beta |Q^-_i><Q^-_i|,\ \ \ \
V_{sep}^{(i+3)}\equiv \beta |Q_{i+3}><Q_{i+3}|=\beta |Q^+_i><Q^+_i|,\ \ \ i=1,2,3.
$$
Let us introduce the definition of the reflection operator~$\Gamma_i$,
following the terminology used in the previous papers,
\be
\Gamma_i= V_{sep}^{(i)} R_i,\ \ \ \ \ R_i=(H_i-E)^{-1},\ \ \ \ \ H_i=H_0+ V_{sep}^{(i)}.
\label{shvarz-def}
\ee
The unperturbed operator~$H_0$ was described in the expression~(\ref{shred-6}).
We also need the definition~$R_0=(H_0-E)^{-1}$.

According to the results obtained in~\cite{BBL-1},
the representation for the resolvent~$R=(H-E)^{-1}$
of the Schr\"odinger operator~$H$~(\ref{shred-6})
can be represented as
\be
R=R_0\left(I-\mathop{\sum}\limits_{i=1}^6\Gamma_i + \mathop{\sum\sum}\limits_{i,j=1,..,6\ i\neq j}\Gamma_i\Gamma_j -
\mathop{\sum\sum\sum}\limits_{i,j,k=1,..,6\ i\neq j,\ j\neq k}\Gamma_i\Gamma_j\Gamma_k-\dots \right).
\label{shvarz-rez}
\ee

In turn, separating the first two terms
in curly brackets in the expression~(\ref{shvarz-rez}),
we obtain
$$
R|Q_b>=(H-E)^{-1}|Q_b>=
$$
$$
=(H_0-E)^{-1}|Q_b>-\mathop{\sum}\limits_{i=1}^6\frac{\beta}{1+\beta F_i(E)}
<Q_i|(H_0-E)^{-1}|Q_b>(H_0-E)^{-1}|Q_i>+
$$
$$
+(H_0-E)^{-1}\left(\mathop{\sum\sum}\limits_{i,j=1,..,6\ i\neq j}\Gamma_i\Gamma_j -
\mathop{\sum\sum\sum}\limits_{i,j,k=1,..,6\ i\neq j,\ j\neq k}\Gamma_i\Gamma_j\Gamma_k-\dots \right).
$$
Here we use the notation
$$
F_i(E)\equiv <Q_i|(H_0-E)^{-1}|Q_i>.
$$

Let us now calculate the first iteration
of the reflection operators~$\Gamma_i\Gamma_j,\ \ i\neq j$
to verify, following the ideology of~\cite{BBL-1},
that the properties of iterations improve
as the order of iteration grows,
\be
(H_0-E)^{-1}\Gamma_i\Gamma_j|Q_b>=
\label{first-it}
 \ee
$$
=\beta^2 (H_0-E)^{-1}|Q_i>\frac{1}{1+\beta F_i(E)} <Q_i|(H_0-E)^{-1}|Q_j>
 \frac{1}{1+\beta F_j(E)} <Q_j|(H_0-E)^{-1}|Q_b>.
$$
Note that the matrix element~$<Q_i|(H_0-E)^{-1}|Q_j>$
admits the following estimate:
\be 
<Q_i|(H_0-E)^{-1}|Q_j> \sim \int_{\Pi_i}dX Q_i(X) \int_{\Pi_j}dX' Q_j(X')\frac{e^{i\sqrt{E}|X-X'|}}{\sqrt{|X-X'|}}=O(R_1^{-1/2}).
\label{small-0}
\ee
Here we used the localization of the functions~$Q_i$ and $Q_j$ (\ref{discrep-ggg})
in the regions~$\Pi_i$ and $\Pi_j$,
\be
\Pi_i\equiv \Pi_i^{-},\ \ \ \ \Pi_{i+3}\equiv \Pi_{i+3}^{+},\ \ \ \ i=1,2,3,
\label{domains}
\ee
respectively, sufficiently remote from each other,
$$
dist\{\Pi_i,\Pi_j\}=O\left(\frac{1}{R_1}\right),\ \ \ \ i\neq j.
$$
We also note that the function~$Q_b$
according to~(\ref{shr-new-ggg}) is localized in the region~$\Pi_1$.
Thus, there arises an~additional smallness
of contributions~(\ref{first-it}) $R_0\Gamma_i\Gamma_j|Q_b>|_{j\neq 1}$,
corresponding to the second iteration of reflection operators,
to the solution of the boundary value problem~(\ref{gran-prob-ggg}),
\be
<Q_j|(H_0-E)^{-1}|Q_b>|_{j\neq 1}=O\left(\frac{1}{\sqrt{R_1}}\right).
\label{small-1}
\ee
Taking into account the estimates~(\ref{small-0}) and~(\ref{small-1}),
we arrive at the conclusion
\be
R_0\Gamma_i\Gamma_j|Q_b>=O\left(\frac{1}{R_1}\right),\ \ \ i\neq j,\ \ j\neq 1.
\label{small-2}
\ee
Note also that according to the estimate~(\ref{first-it}),
summation over the index~$i,\ \ i\neq j$
in the expression~$\Gamma_i\Gamma_j|Q_b>$
leads to a~``projection'' of the bounded
(taking into account the boundary conditions)
function~$(H_0-E)^{-1}Q_b$ by the operator
$$
\tilde{P}_{ij}\equiv \mathop{\sum}\limits_{i\neq j}|Q_i><Q_i|
$$
onto the region of support
of the corresponding residual function.
We take into account in this case
that the spectral function~$F_i(E)$
in this problem does not depend
on the value of the index,
and we also take into account the boundedness
of the norm~$||Q_i||_{L_2}$. Thus, the additional summation
does not lead to an~increase
of the corresponding iteration of reflection operators.

We note that direct calculations show
that the order of the~$m$th iteration
of the reflection operators is defined as~$O(R_1^{-m/2})$.

We can now conclude that the solution
of the boundary value problem~(\ref{gran-prob-ggg})
with a~sufficiently large value of the parameter~$R_1$
and some value of~$R_2$, $\ R_1\ll R_2\ll R$,
can be described as
\be
\Phi=R|Q_b>=(H_0-E)^{-1}|Q_b>-
\label{phi-sol}
\ee
$$
-\beta\mathop{\sum}\limits_{i=1}^6 (H_0-E)^{-1}|Q_i>\frac{1}{1+\beta F_i(E)}
<Q_i|(H_0-E)^{-1}|Q_b>+
$$
$$
+\beta^2\mathop{\sum\sum}\limits_{i,j=1,..,6\ i\neq j}
 (H_0-E)^{-1}|Q_i>\frac{1}{1+\beta F_i(E)} <Q_i|(H_0-E)^{-1}|Q_j>
 \frac{1}{1+\beta F_j(E)} <Q_j|(H_0-E)^{-1}|Q_b>+
$$
$$
+O\left(\frac{1}{R^{3/2}_1}\right).
$$
This expression is defined by solving a~set of boundary value problems
in a~circle of large radius~$R$
for the operator~$H_0$ of the form
$$
\left\{
\begin{tabular}{l}
$(H_0-E)\tilde{\Phi}=\tilde{Q}$ \\
$\left(\frac{\partial \tilde{\Phi}}{\partial n} - i\sqrt{E}\tilde{\Phi}\right)|_{\partial B_R}=
O\left(\frac{1}{R^{3/2}}\right)$\\
\end{tabular}
\right.
$$
for various values of the right-hand side,
$$
\tilde{Q}=Q_b,\ Q_i,\ i=1,2,\dots,6.
$$

We note that with an~increase in the kinetic energy of the cluster
relative to the third particle~$p^2$ and, thereby, with a~decrease
in the parameter~$\beta$ (\ref{beta-def}),
the representation for the solution~(\ref{phi-sol})
of the boundary value problem is simplified.

The solution~$\Phi$ described in~(\ref{phi-sol})
after substitution into the expressions~(\ref{koeff-ggg}) and~(\ref{koeff-in-ggg})
defines the values of cluster amplitudes~$a_j^{\ tau_j},\ \ j=1,2,3;\ \tau_j=\pm$ of the processes~$2\rightarrow 2$.
The three-body scattering amplitude~$A(\theta)$
of the~$2\rightarrow 3$ process is defined by the expression
$$
A(\theta)=\Phi(\bX)\sqrt{X}\exp\{-i\sqrt{E}X\}|_{X=R}+O\left(\frac{1}{R}\right) .
$$
In terms of the function~$\Phi$ (\ref{phi-sol}),
the expression~(\ref{exact-ggg})
defines the complete asymptotic solution
of the~$2\rightarrow 2(3)$ scattering problem.

\section{Numerical analysis}

Let us represent the results of the numerical analysis
carried out according to the outline proposed above.
The calculations were carried out
using the FreeFem++ computer package.
The function~$\Phi$ was calculated up to and including the second iteration
of the reflection operators
according to the expression (\ref{phi-sol})
for the total energy of the system~$E=0.5$,~$E=1$, and~$E=2$.
The pair potential~$v(x)$ is chosen as follows:
$$
v(x)=\left\{
\begin{tabular}{l}
$0,\ \ x\in (-\infty,-1)\cap (1,+\infty)$,\\
$-1\ \ x\in (-1,1)$.\\
\end{tabular}
\right.
$$
This potential supports a~unique bound state~$\varepsilon=-0.453753$.

The result of calculating the absolute value
of the function~$\Phi$ in the case~$E=0.5$
is represented in Fig.~1.

\includegraphics[scale=0.5]{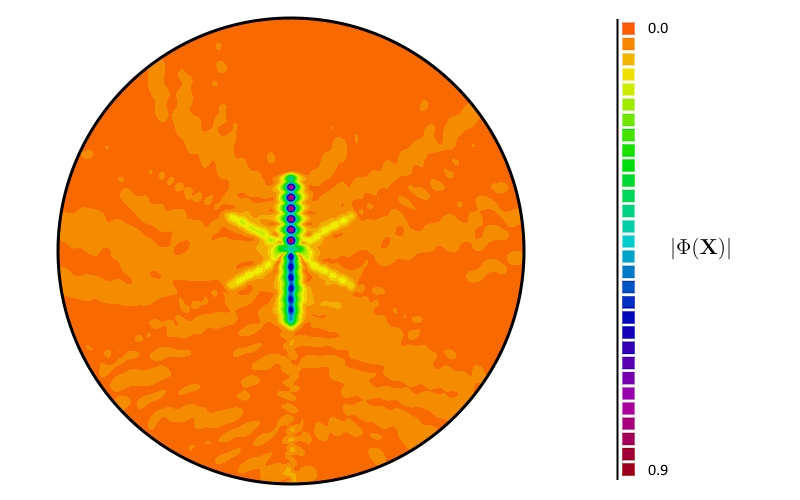}

\parbox[t]{5cm}{Fig.~1}

In the figure, screen~$l_1$ is directed vertically
from top to bottom.
Screens~$l_2$ and~$l_3$ are rotated
relative to~$l_1$ by~$2\pi/3$ and~$4\pi/3$, respectively.
Thus, the incident wave moves
towards the center of the circle
along the vertical axis from top to bottom.

According to formulas~(\ref{koeff-ggg}) and~(\ref{koeff-in-ggg}),
the two-particle cluster amplitudes~$a_j^{\tau_j},\ \ j=1,2,3,\ \ \tau_j=\pm$ were calculated.
We represent the obtained values of the two-particle scattering amplitudes,
$$
a_1^-=0.35771+0.0302174i,\ \ a_2^+= -0.176662+0.0142423i,\ \ a_3^-=-0.144211+0.00353035i,\ \
$$
$$
a_1^+= -0.194684+0.861316i,\ \ a_2^-=-0.147772-0.00801603i,\ \ a_3^+= -0.143487+0.0180234i.
$$
It is clear that the probability
of a~cluster wave passing through without rearrangement
prevails, which corresponds to the coefficient~$a_1^+$.

The absolute value of the amplitude of the breakup~$A(\theta)$
of the system into three particles for the energy~$E=0.5$ is represented in Fig.~2.

\includegraphics[scale=0.5]{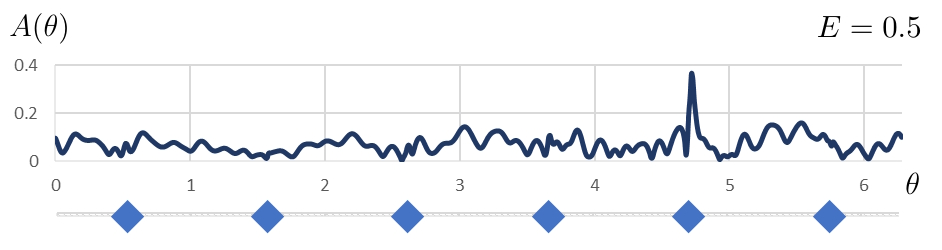}

\parbox[t]{5cm}{Fig.~2}

Blue diamonds show the angular position of the ``half-screens'' on the coordinate plane.
The only significant peak falls on the ``half-screen''~$l_1^+$
and corresponds to the maximum of the breakup amplitude
in the vicinity of the forward scattering direction.

We also note that according to the conservation law of total probability,
the calculated total probability of all processes
admissible in the system with high accuracy is equal to 1,
$$
\mathop{\sum}\limits_{j=1}^{3}\mathop{\sum}\limits_{\tau_j=\pm}|a_j^{\tau_j}|^2 +
\int_0^{2\pi} |A(\theta)|^2 d\theta= 1.04.
$$

The calculation of cluster amplitudes~$a_j^{\tau_j},\ \ j=1,2,3;\ \tau_j=\pm$
allowed, in turn, calculating according to~(\ref{exact-ggg})
the expression~$\ Psi_0=\Psi-\Phi$, i.e.,
a~family of cluster contributions to the solution
(taking into account the incident wave),
multiplied by the radial cut-off function,
$$
\Psi_0(\bX,\bP) =  \psi_{in}(\bX,\bP)\zeta(X) +
\mathop{\sum}\limits_{j=1}^3
\mathop{\sum}\limits_{\tau_j\in\{+,-\}}
a_j^{\tau_j}\psi_j^{\tau_j}(\bX,\bP)\zeta(X).
$$
Figure~3 shows the calculation results for~$|\Psi_0|$ for the case $E=0.5$

\includegraphics[scale=0.5]{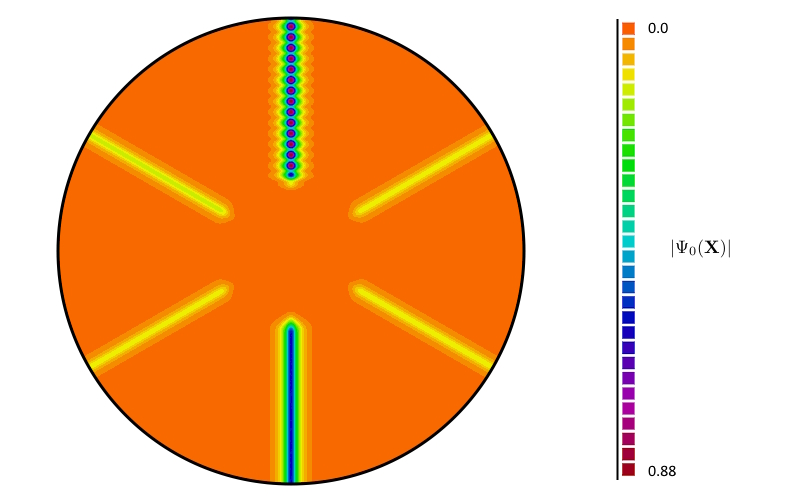}

\parbox[t]{5cm}{Fig.~3}

Finally, the complete solution of the scattering problem~$\Psi$
(its absolute value) for~$E=0.5$ is represented in Fig.~4

\includegraphics[scale=0.5]{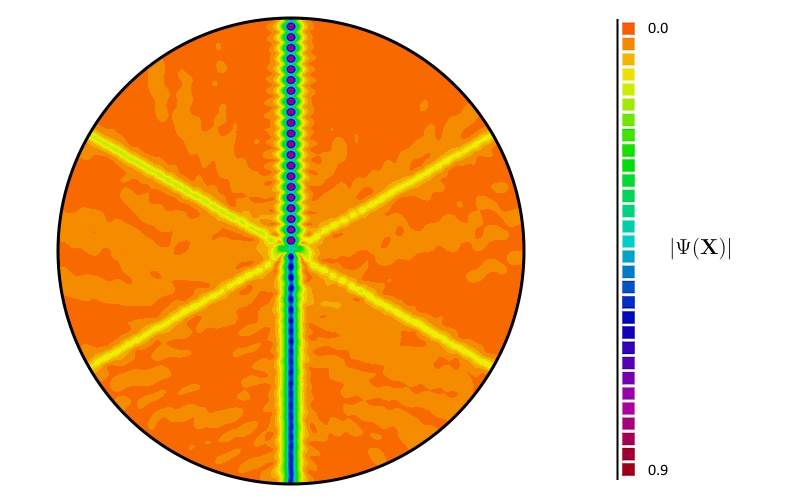}

\parbox[t]{5cm}{Fig.~4}

Figures~5--7 show the results of similar calculations for~$E=2$

\includegraphics[scale=0.5]{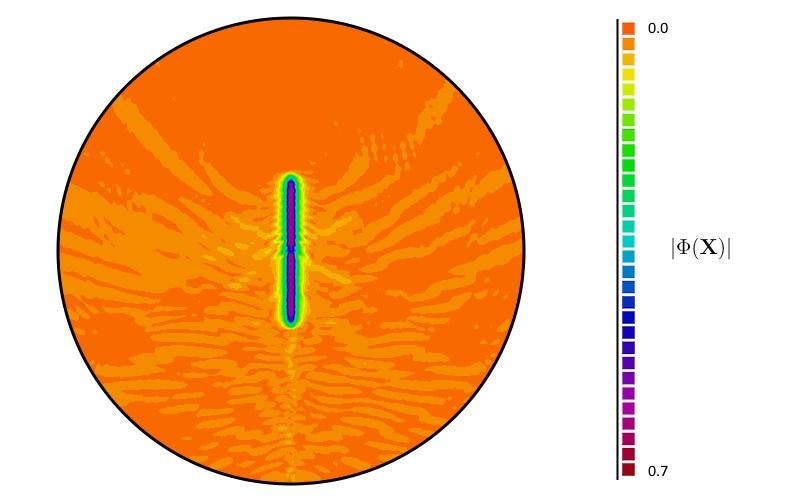}

\parbox[t]{5cm}{Fig.~5}

\vskip1cm

\includegraphics[scale=0.5]{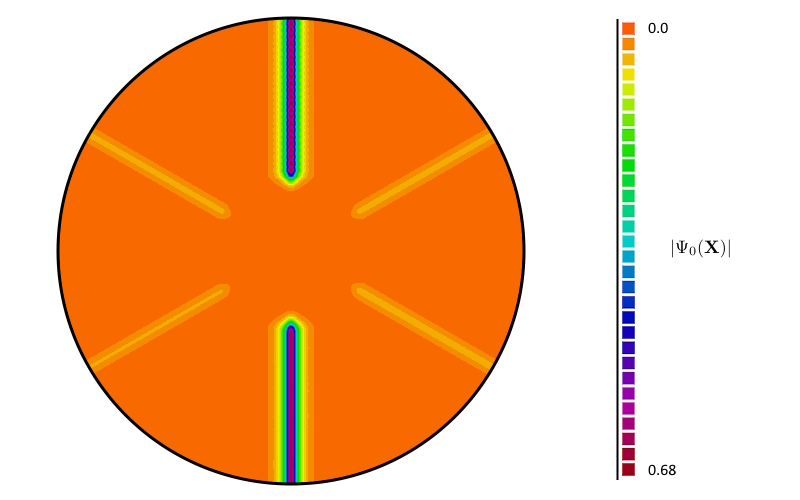}

\parbox[t]{5cm}{Fig.~6}

\vskip1cm

\includegraphics[scale=0.5]{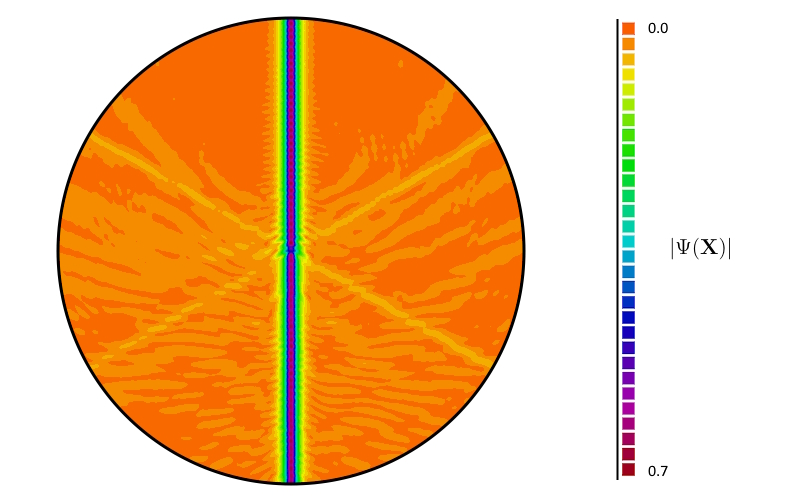}

\parbox[t]{5cm}{Fig.~7}

We also represent the values of cluster amplitudes for the case~$E=2$,
$$
a_1^-=0.0414651-0.0364133i,\ \ a_2^+=0.0493592-0.0132862i,\ \ a_3^-=-0.0403836+0.0184125i,
$$
$$
a_1^+=0.285251+0.9518i,\ \ a_2^-=-0.0541078+0.0147765i,\ \ a_3^+=0.0518601-0.00337697i.
$$

Comparing Figs.~4 and~7, we note that, as expected,
the influence of reclustering channels (the cluster rearrangement)
increases at lower energy. The amplitudes~$a_j^{\tau_j},\ \ j\neq 1$
at lower energies increase in magnitude.
This is also clear from the given values
of the coefficients themselves.

\section{Interpretation of the results obtained and the conclusion}

We note that the method proposed in this paper for searching
for an~asymptotic solution of the quantum scattering problem~$2\rightarrow 2(3)$
is based on the diffraction approach to the scattering problem.
The ideas of this approach were proposed in~\cite{BMS1}--\cite{BMS2}
and later developed in~\cite{BL-1} for the case of three
one-dimensional quantum particles
with finite repulsive pair potentials.
In this paper,
the case of attractive pair potentials
supporting the bound states in pair subsystems
is consistently considered for the first time
within the framework of the diffraction approach. The paper considers
the case of~$2\rightarrow 2(3)$ scattering
(the scattering of a~bound pair on a~third particle),
however, the described outline can be also easily generalized
to the~$3\rightarrow 2(3)$ scattering problem
(all three particles are free in the initial state).

Both in this problem and in the problems considered in~\cite{BMS1}--\cite{BL-1},
after separating the center of mass of the system of three particles,
the coordinate space is a~plane. The support of the total potential
on this plane is the system of three infinite semitransparent strips
intersecting in a~compact region.
Each such strip with index~$j,\ j=1,2,3$ includes
two ``half-screens''~$\ l_j^{\tau_j},\ \tau_j=\pm$,
described in Sec.~2.
This system of strips divides the coordinate plane into six sectors.
Each sector corresponds to a~region of the coordinate space
where a~certain permutation of three particles
on the axis is fixed. At the same time, passing through the~``screen''
and penetrating into the neighboring sector corresponds
to the tunneling of a~pair of particles
through each other while the position of the third particle remains unchanged.

The diffraction outline corresponding to the~$3\rightarrow 3$ problem,
described in detail in~\cite{BMS1}--\cite{BL-1},
as well as numerically studied in~\cite{BL2}--\cite{BL3},
corresponds to the scattering of the plane wave (three free particles)
coming from infinity in an~opening of one fixed sector
(corresponding to the initial permutation of particles),
on the system of three such infinite semitransparent strips.
A~detailed analysis of such a~problem was made
in the papers mentioned above.

The diffraction outline corresponding
to the~$2\rightarrow 2(3)$ scattering problem
and considered in this paper is somewhat different
from the one mentioned above.
The localized (with respect to the variable~$x_1$) state
(the first term~$\psi_{in}$ in the representation~(\ref{ass-ggg}))
with moment~$p$ is scattered in the central region,
that is, the vicinity of the intersection of three supports of potential.
As a result of scattering, localized states
moving away from the central region with amplitudes~$a_j^{\tau_j}$
along the ``half-screens''~$\ l_j^{\tau_j}$ are excited
provided that the pair potential~$v_j$ supports the bound states.
These waves are described by the second term
in the representation~(\ref{ass-ggg}) and describe the processes
of rearrangement of pair subsystems. In other words,
the family of such contributions to the solution
of the scattering problem corresponds to the~$2\rightarrow 2$ scattering channel.
Another result of scattering
of the localized state~$\psi_{in}$ on the central region
is the diverging circular wave with amplitude~$A(\hX,\bP)$,
which is the third term in the asymptotics~(\ref{ass-ggg}).
This term corresponds to the~$2\rightarrow 3$ scattering channel,
that is, to the channel of the system’s breakup into three free particles.

The main problem in describing such scattering processes
is to define the amplitudes of the rearrangement~$a_j^{\tau_j}$
and the amplitudes of the breakup~$A(\hX,\bP)$.
Moreover, the problem of finding the amplitudes~$a_j^{\tau_j}$
is as complicated as the one of finding~$A(\hX,\bP)$
since the formation of all amplitudes is defined
not by the asymptotic region, but by the compact vicinity
of the intersection point of the three ``screens''.
We solve this problem in two stages. At the first stage,
we exclude the cluster channels from consideration, obtaining in return
an~additional separable potential~(\ref{sep-sh}),
$$
V_{sep}=\mathop{\sum}\limits_{i=1}^6 V_{sep}^{(i)},\ \ \ \ \ \  V_{sep}^{(i)}\equiv \beta |Q_i><Q_i|,
$$
in the Hamiltonian operator. The support of this separable potential
is localized in the regions~$\Pi_i,\ \ i=1,2,\dots,6$
of conjugation of the cluster channels
and the region of interchannel interaction.
In each of the regions~$\Pi_i$ (\ref{domains}),
the potential is defined
by the corresponding residual function~$Q_i$
of the distorted cluster solution in the Schr\"odinger equation.
After eliminating the cluster channels, a~boundary value problem
is constructed to solve~$\Phi$ in a~three-particle channel.

The second stage is related to the solution
of this boundary value problem. Relation (\ref{phi-sol})
obtained for the function~$\Phi$
admits the diffraction interpretation in the following sense.
The diverging circular wave generated by the source~$Q_b$
localized in the region~$\Pi_1$, in turn,
is scattered on all sources~$Q_i,\ \ i\neq 1$,
acquiring in this case an~amplitude~$\frac{\beta}{1 +\beta F_i(E)}<Q_i|(H_0-E)^{-1}|Q_b>$.
Each of the waves scattered by the source~$Q_i$
is scattered by the source~$Q_j,\ \ j\neq i$,
acquiring an~amplitude~$\frac{\beta}{1+\beta F_j(E)}<Q_j|(H_0-E)^{ -1}|Q_i>$.
The family of contributions from such rescatterings
is generated by the interaction between a two-particle (cluster)
and three-particle channel (the breakup channel)
and forms the solution of the boundary value problem~$\Phi$
in the three-particle channel. In turn,
the scattering amplitudes in the cluster channel
are defined as functionals of~$\Phi$
according to~(\ref{koeff-ggg}) and~(\ref{koeff-in-ggg}).

In conclusion, we note that the cut-off function
introduced in the expression~(\ref{srez-ggg})
and formally defining the region of interaction
between the channels of different types does not introduce,
as it may seem at first glance,
arbitrariness into the solution of the complete problem.
The results of numerical calculations
represented in the previous section
at total energy~$E=0.5$ and~$E=2$ clearly demonstrate
that although the region of variation in the cut-off function
is clearly observed, for example, in Figs.~1 and~3, it is absent in Fig.~4.
A~similar effect is observed in Figs.~5,~6, and~7.
This means that the computational outline
proposed in this paper does not depend at the final stage
on the influence of the cut-off function,
which plays such a~significant role
at the intermediate stages of calculations.

\section{Funding}

This work was financially supported by the Russian Science
Foundation, project no.~22-11-00046.

\end{document}